# Genetic Algorithm Based Improved Sub-Optimal Model Reduction in Nyquist Plane for Optimal Tuning Rule Extraction of PID and $PI^\lambda D^\mu$ Controllers via Genetic Programming

Saptarshi Das, Indranil Pan, Shantanu Das, and Amitava Gupta

*Abstract*—Genetic Algorithm (GA) has been used in this paper for a new Nyquist based sub-optimal model reduction and optimal time domain tuning of PID and fractional order (FO) $PI^\lambda D^\mu$ controllers. Comparative studies show that the new model reduction technique outperforms the conventional $H_2$-norm based reduced order modeling techniques. Optimum tuning rule has been developed next with a test-bench of higher order processes via Genetic Programming (GP) with minimum value of weighted integral error index and control signal. From the Pareto optimal front which is a trade-off between the complexity of the formulae and control performance, an efficient set of tuning rules has been generated for time domain optimal PID and $PI^\lambda D^\mu$ controllers.

## I. INTRODUCTION

EMPIRICAL rules are classically used to tune PID controllers and are very popular in process control since the advent of PID controllers. These rules are mainly devised from certain design specification in time or frequency domain. O' Dwyer [1] has tabulated several optimal PI/PID controller tuning rules for various types of reduced order processes based on diverse control objectives like set-point tracking, load disturbance rejection etc. The conventional step-response process reaction curve based graphical method to obtain First Order Plus Time Delay (FOPTD) models for unknown processes has been extended by Skogestad in [2] for PID controller tuning. Performance comparison of well established empirical rules like Ziegler-Nichols (Z-N), refined Ziegler-Nichols, Cohen-Coon (C-C), Internal Model Control (IMC), Gain-Phase Margin (GPM) have been studied by Tan *et al.* [3] and Lin *et al.* [4]. Also, Ho *et al.* done a comparative study for integral performance indices based optimum parameter settings for PI controller in [5] and PID controller in [6]. Impact of choosing different performance index like Integral of Time Multiplied Absolute Error (ITAE) or Integral of Time Multiplied Squared Error (ITSE) corresponding to set-point tracking and load rejection on the optimum tuning formula has been studied by Zhuang & Atherton [7]. The idea has been extended in Mann *et al.* [8] considering actuator constraints. Ho *et al.* [9] combined the concept of time domain performance index optimization and gain-phase margin or GPM method to develop improved tuning rules.

It is well known that for the development of tuning formula for an identified process, it needs to be reduced in a suitable template like FOPTD or Second Order Plus Time Delay (SOPTD) etc, since these rules are basically a mapping between the process and optimum controller parameters. Zhuang & Atherton [7] proposed the tuning formula for PID controllers to handle FOPTD processes, which is rather poor approximation for higher order processes as shown by Astrom & Hagglund [10]. Zhuang & Atherton [7] used several higher moments of time and error terms in the integral performance index which puts higher penalties for larger error and sluggish response, yielding large control signal which may saturate the actuator. This paper tries to extend the idea for the tuning of PID and FOPID controllers while taking the integral of error index and the control signal together which is viewed like a trade-off between the ability of set-point tracking and required cost of control [11]. The optimum time domain tuning of PID type controllers are attempted with genetic algorithm as studied with similar objectives [11]-[13]. These optimal integral performance indices based tuning methods for PID controllers show nice closed loop behavior in terms of low overshoot and settling time but the only requirement is that the process model has to be identified accurately. For higher order process models simple FOPTD reduced order approximations give larger modeling errors which may produce inferior closed loop response with the available tuning rules. Hence, an improved sub-optimal model reduction in the Nyquist plane is attempted first to reduce higher order processes in SOPTD template which is a better approximation than the FOPTD [10]. Reduction in SOPTD template for improved frequency domain tuning of PID controllers has been extensively studied by Wang *et al.* [14].

Also, Zhuang & Atherton [7] developed the optimum tuning formula based on the least-square curve fitting technique with the tested optimum controller parameters with few FOPTD models. Such a chosen structure based linear fitting method indeed reduces the accuracy of the tuning formula which is further enhanced in this paper with a much sophisticated technique i.e. a Genetic Programming based approach. Contemporary researchers like Valerio & Sa da Costa [15], Chen *et al.* [16], Padula & Visioli [17] developed tuning rules for FOPID controllers but the idea of this paper is to extract the rules in an optimal fashion via GP with GA based optimum reduced parameter model and

Manuscript received April 14, 2011. This work has been supported by the Department of Science & Technology (DST), Govt. of India under the PURSE programme.

S. Das is with School of Nuclear Studies and Applications (SNSA), Jadavpur University, Salt-Lake Campus, LB-8, Sector 3, Kolkata-700098, India. (E-mail: saptarshi@pe.jusl.ac.in).
I. Pan and A. Gupta are with Dept. of Power Engineering, Jadavpur University, Salt-Lake Campus, LB-8, Sector 3, Kolkata-700098, India.
Sh. Das is with Reactor Control Division, Bhabha Atomic Research Centre, Mumbai-400085, India.

PID/FOPID parameters. The rationale behind using Genetic Programming is the fact that it is based on symbolic regression which searches for not only the optimal parameters within a structure but also the structure itself i.e. the optimal PID/FOPID controller tuning formulae in our case that ensures low error index and control signal.

The rest of the paper is organized as follows. Section II discusses about a new sub-optimal model reduction for higher order processes. Section III shows the GA based optimal PID/FOPID controller tuning results and GP based tuning rule generation with the achievable closed loop performances. The paper ends with conclusion in section IV, followed by the references.

## II. NEW APPROACH OF SUB-OPTIMAL MODEL REDUCTION

### A. New Optimization Framework

Xue & Chen [18] proposed a novel method of reducing higher order process models by minimizing the $H_2$ norm of the original higher order model $P(s)$ and reduced order model $\tilde{P}(s)$ using an unconstrained optimization. i.e.

$$J_{2-norm} = \|P(s) - \tilde{P}(s)\|_2 \tag{1}$$

where, $\|\cdot\|_2$ denotes the 2-norm of a system which is a measure of the energy of a stable LTI system with an impulse excitation and is given by the following expression:

$$\|P(s)\|_2 = \sqrt{\frac{1}{2\pi} \int_{-\infty}^{\infty} trace\left[P(j\omega)\overline{P(j\omega)^T}\right] d\omega} \tag{2}$$

In this paper we have used another optimization framework which minimizes the discrepancy between the frequency responses of the higher order and reduced parameter process model in the complex Nyquist plane. The proposed methodology has been found to produce better accuracy in the model reduction process, since the $H_2$ norm based method, discussed earlier [18] is based on the minimization of discrepancy in the magnitude curves only. The proposed Nyquist based method minimizes both the discrepancies in the gain and phase of the two said systems. The proposed objective function for model reduction is given by (3):

$$J_{nyquist} = w_1 \cdot \left\|\text{Re}\left[P(j\omega)\right] - \text{Re}\left[\tilde{P}(j\omega)\right]\right\| + w_2 \cdot \left\|\text{Im}\left[P(j\omega)\right] - \text{Im}\left[\tilde{P}(j\omega)\right]\right\| \tag{3}$$

Here, the norm $\|\cdot\|$ denotes Euclidian length of the vectors. The weights $\{w_1, w_2\}$ are chosen to be equal so as not to emphasize discrepancies either in the real or imaginary part of the frequency response. To evaluate the objective function (3) in each iterations, within an optimization framework, logarithmically spaced 500 frequency points have been taken within the frequency-band of $\omega \in [\omega_l, \omega_h] = [10^{-4}, 10^4] Hz$. Here, the two objective functions (1) and (3) denotes the discrepancies in the $H_2$ norm and the real and imaginary parts of the Nyquist curves corresponding to the higher order and reduced order models. The new objective function (3) is now minimized with an unconstrained Genetic Algorithm to obtain the reduced parameter models in a FOPTD (5) as well as SOPTD (6) templates with the corresponding sub-optimal reduced order parameters in Table 1 for a test-bench of higher order processes. The model reduction technique has been termed as "sub-optimal" due to the fact that it extracts the apparent delays ($L$) in the higher order models with an equivalent third order Pade approximation:

$$e^{-Ls} \simeq \frac{-L^3 s^3 + 12L^2 s^2 - 60L + 120}{L^3 s^3 + 12L^2 s^2 + 60L + 120} \tag{4}$$

Here, the reduced order templates are given as:

$$P_{FOPTD}(s) = \frac{Ke^{-Ls}}{(\tau s + 1)} \tag{5}$$

$$P_{SOPTD}(s) = \frac{Ke^{-Ls}}{(\tau_{max} s + 1)(\tau_{min} s + 1)} \tag{6}$$

with the reduced order parameters $\{K, \tau, L\}$ denoting the dc-gain, time-constant (maximum or minimum) and time-delay respectively.

### B. Test-Bench Processes

In this paper, four set of higher order test bench processes (7)-(10) have been studied as reported by Astrom & Hagglund [19]. $P_1$ represents a class of higher order processes with concurrent poles. $P_2$ represents a class of fourth order processes with increasing order of smallest time constants ($\alpha$). $P_3$ represents a class of third order processes with different values of the repeated dominant/non-dominant time constant ($T$). $P_4$ represents a class of non-minimum phase processes with increasing magnitude of the real right half plane zero.

$$P_1(s) = \frac{1}{(1+s)^n}, n \in \{3, 4, 5, 6, 7, 8, 10, 20\} \tag{7}$$

$$P_2(s) = \frac{1}{(1+s)(1+\alpha s)(1+\alpha^2 s)(1+\alpha^3 s)}, \tag{8}$$

$$\alpha \in \{0.1, 0.2, 0.3, 0.4, 0.5, 0.6, 0.7, 0.8, 0.9\}$$

$$P_3(s) = \frac{1}{(1+s)(1+sT)^2}, \tag{9}$$

$$T \in \{0.005, 0.01, 0.02, 0.05, 0.1, 0.2, 0.5, 2, 5, 10\}$$

$$P_4(s) = \frac{(1-\alpha s)}{(1+s)^3}, \tag{10}$$

$$\alpha = \{0.1, 0.2, 0.3, 0.4, 0.5, 0.6, 0.7, 0.8, 0.9, 1.0, 1.1\}$$

The accuracies of the GA based optimization for model reduction using $H_2$ norm based and proposed Nyquist based approach has been compared in Fig. 1-4. It is clear that the proposed model reduction technique produces SOPTD models with high degree of accuracy in the Nyquist plane. Also, FOPTD models for the test processes are less accurate than SOPTD with modeling objectives (1) and (3).

TABLE I
NYQUIST BASED PROPOSED MODEL REDUCTION RESULTS

| Class of Processes | Varying Parameter | Sub-optimum SOPTD Parameters | | | | |
|---|---|---|---|---|---|---|
| | | $J_{min}$ | K | $\tau_{max}$ | $\tau_{min}$ | L |
| $P_1$ | n=3 | 0.35763 | 1.0 | 1.335035 | 1.296596 | 0.458524 |
| | n=4 | 0.534457 | 1.0 | 1.586542 | 1.548473 | 1.03317 |
| | n=5 | 0.643986 | 1.0 | 1.797635 | 1.770904 | 1.666146 |
| | n=6 | 0.720594 | 1.0 | 1.989875 | 1.959647 | 2.344943 |
| | n=7 | 0.779376 | 1.0 | 2.163055 | 2.14323 | 3.051016 |
| | n=8 | 0.82832 | 1.0 | 2.310304 | 2.310215 | 3.782639 |
| | n=10 | 0.91604 | 1.0 | 2.661457 | 2.549809 | 5.293009 |
| | n=20 | 2.504335 | 1.0 | 5.451683 | 5.397813 | 9.999728 |
| $P_2$ | α=0.1 | 0.004308 | 1.0 | 0.999772 | 0.100915 | 0.010279 |
| | α=0.2 | 0.028107 | 1.0 | 0.992451 | 0.214076 | 0.038794 |
| | α=0.3 | 0.060572 | 1.0 | 0.979505 | 0.341498 | 0.092874 |
| | α=0.4 | 0.107937 | 1.0 | 0.943464 | 0.51063 | 0.167586 |
| | α=0.5 | 0.173435 | 1.0 | 0.833884 | 0.778235 | 0.270018 |
| | α=0.6 | 0.292888 | 1.0 | 0.919789 | 0.886179 | 0.409777 |
| | α=0.7 | 0.400586 | 1.0 | 1.026115 | 1.021073 | 0.559864 |
| | α=0.8 | 0.480812 | 1.0 | 1.233382 | 1.10547 | 0.720248 |
| | α=0.9 | 0.521566 | 1.0 | 1.371358 | 1.331686 | 0.879882 |
| $P_3$ | T=0.005 | 0.003451 | 1.0 | 1.000027 | 0.007301 | 0.00276 |
| | T=0.01 | 0.006693 | 1.0 | 0.999721 | 0.014931 | 0.005228 |
| | T=0.02 | 0.013254 | 1.0 | 0.999557 | 0.030272 | 0.010203 |
| | T=0.05 | 0.031173 | 1.0 | 0.997605 | 0.075538 | 0.026398 |
| | T=0.1 | 0.05823 | 1.0 | 0.989257 | 0.157307 | 0.050227 |
| | T=0.2 | 0.100513 | 1.0 | 0.963887 | 0.337572 | 0.09348 |
| | T=0.5 | 0.243507 | 1.0 | 0.911085 | 0.868222 | 0.253221 |
| | T=2 | 0.274858 | 1.0 | 2.285902 | 2.162089 | 0.662506 |
| | T=5 | 0.105979 | 1.0 | 5.271248 | 4.954549 | 0.85439 |
| | T=10 | 0.048469 | 1.0 | 9.999702 | 9.998882 | 0.98878 |
| $P_4$ | α=0.1 | 0.350007 | 1.0 | 1.321307 | 1.304839 | 0.562264 |
| | α=0.2 | 0.334032 | 1.0 | 1.317905 | 1.293675 | 0.66746 |
| | α=0.3 | 0.332085 | 1.0 | 1.393695 | 1.197571 | 0.773718 |
| | α=0.4 | 0.351824 | 1.0 | 1.334063 | 1.234247 | 0.873208 |
| | α=0.5 | 0.423653 | 1.0 | 1.298311 | 1.242496 | 0.968798 |
| | α=0.6 | 0.542731 | 1.0 | 1.25362 | 1.252805 | 1.064005 |
| | α=0.7 | 0.698068 | 1.0 | 1.241163 | 1.240979 | 1.150465 |
| | α=0.8 | 0.881815 | 1.0 | 1.293128 | 1.161037 | 1.234179 |
| | α=0.9 | 1.085803 | 1.0 | 1.28306 | 1.138877 | 1.308246 |
| | α=1.0 | 1.307159 | 1.0 | 1.298524 | 1.09749 | 1.387555 |
| | α=1.1 | 1.542905 | 1.0 | 1.312971 | 1.053957 | 1.459166 |

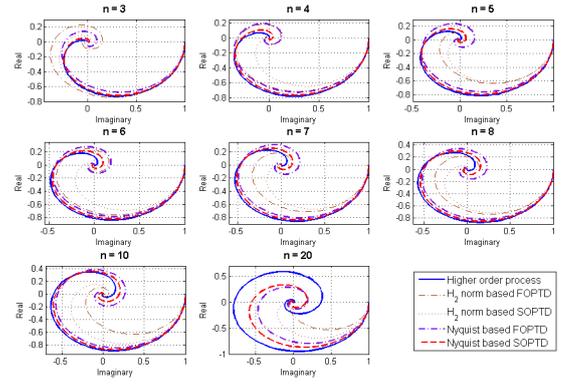

Fig. 1. Accuracies of reduced parameter models of $P_1$ in the Nyquist plane.

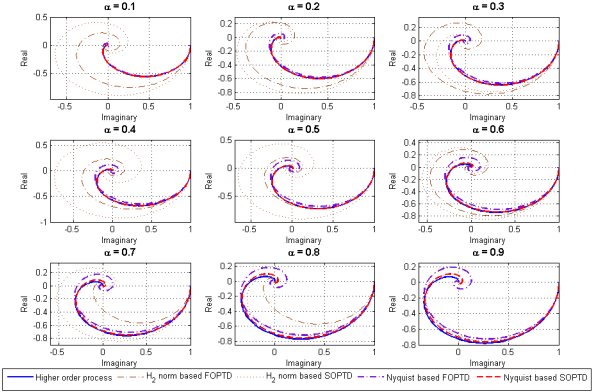

Fig. 2. Accuracies of reduced parameter models of $P_2$ in the Nyquist plane.

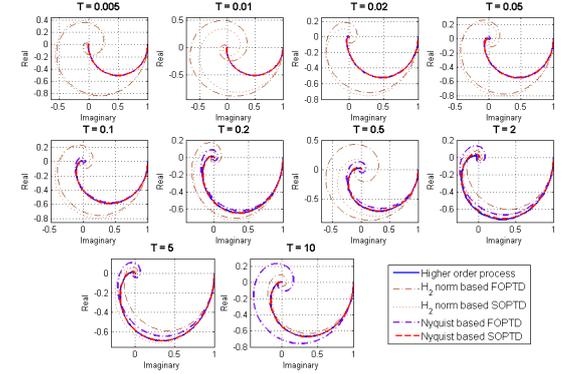

Fig. 3. Accuracies of reduced parameter models of $P_3$ in the Nyquist plane.

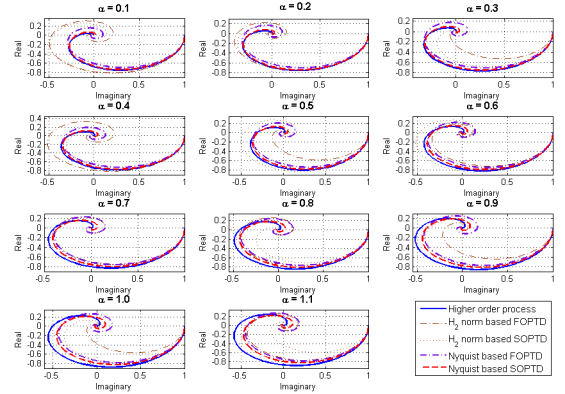

Fig. 4. Accuracies of reduced parameter models of $P_4$ in the Nyquist plane.

## III. TIME DOMAIN OPTIMAL CONTROLLER TUNING

### A. Controller Structures and Their Optimal Tuning

In this paper, the performance of two classes of controllers has been studied to control few higher order processes (7)-(10). The chosen controllers are conventional PID type which is widely used in process control industries and its analogous fractional order $PI^\lambda D^\mu$, proposed by Podlubny [20] which is gaining increased interest amongst the research community. The $PI^\lambda D^\mu$ controller has been considered to have a parallel structure (11) similar to the conventional PID controller [1].

$$C_{FOPID}(s) = K_p + \frac{K_i}{s^\lambda} + K_d s^\mu \qquad (11)$$

Clearly, the $PI^\lambda D^\mu$ controller (11) is a generalization of the classical PID controller with two extra tuning knob i.e. the differ-integral orders $\{\lambda, \mu\}$. The conventional PID controller can be designed with the same technique by putting $\{\lambda, \mu\} = 1$. The PID and $PI^\lambda D^\mu$ controllers are now tuned with a constrained Genetic Algorithm, since its unconstrained version may produce large controller gains and increase the cost of hardware implementation. The goal of the constrained optimization is to minimize a weighted sum of a suitable error index and the control signal (12) similar to that in Pan et al. [11]:

$$J = \int_0^\infty \left[ w_1 \cdot t \cdot |e(t)| + w_2 \cdot u^2(t) \right] \qquad (12)$$

Here, the first term corresponds to the ITAE which minimizes the overshoot and settling time, whereas the second term denotes the Integral of Squared Controller Output (ISCO). The two weights $\{w_1, w_2\}$ balances the impact of control loop error (oscillation and/or sluggishness) and control signal (larger actuator size and chance of integral wind-up) and both have been chosen to be unity in the present simulation study indicating same penalty for large magnitude of ITAE and ISCO.

### B. Genetic Programming (GP) Based Rule Extraction for Optimal Tuning of PID/$PI^\lambda D^\mu$ Controllers

The $PI^\lambda D^\mu$ controller parameters $\{K_p, K_i, K_d, \lambda, \mu\}$ are tuned next with GA minimizing the objective function (12) for the higher order test-bench processes (7)-(10). In this paper, Genetic Programming is used to optimally map the enhanced sub-optimal reduced order SOPTD parameters representing the higher order systems (in Table I) and the optimal PID/$PI^\lambda D^\mu$ controller parameters for extracting optimal tuning rules based on the control objective (12) in an optimum fashion. For tuning rule development several measures of standard SOPTD templates like time-delay ($L$), maximum-minimum time-constant ratio ($\tau_{max}/\tau_{min}$), time-delay to time constant ratio ($L/\tau_{min}$ and $L/\tau_{max}$) etc have been used to map the GA based sub-optimal reduced order SOPTD model parameters with GA based PID/FOPID parameters. O' Dwyer [1] reported least square based empirical rule extraction approach to fit a chosen structure of the tuning rule. The idea has been significantly improved with a GP based approach with optimal choice of the structure to fit the data in the rule and with additional choice of the complexity, representing the tuning formula.

GP [21] is a class of computational intelligence techniques which extends the notion of the conventional Genetic Algorithm, to evolve computer programs which can perform user defined tasks. It is an evolutionary algorithm and is based on the biological strategies of reproduction, crossover and mutation to evolve fitter solutions in the future generations. In the present paper, GP is used for symbolic regression to find out an analytic expression that maps the input variables of the process parameters to the output values of the controller parameters while minimizing the mean absolute error (MAE) of the predicted controller parameters (from the rule) and the specified well-tuned values. Thus instead of finding the coefficients of a particular structure as in the conventional regression in [1], [7] GP searches in the infinite dimensional functional space to find an optimum structure along with the numerical coefficients, minimizing MAE of the controller parameters.

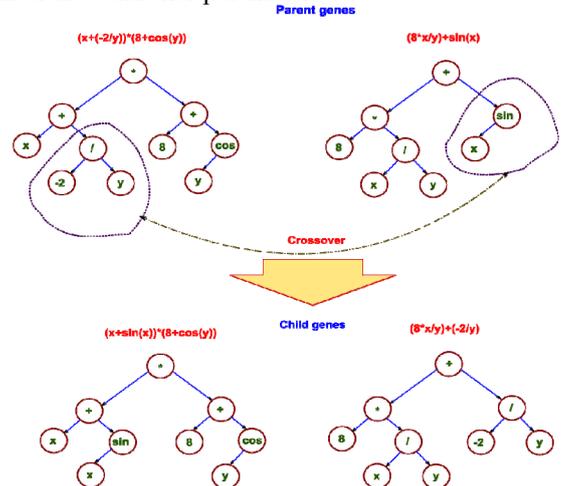
Fig. 5. Schematic of cross-over in Genetic Programming.

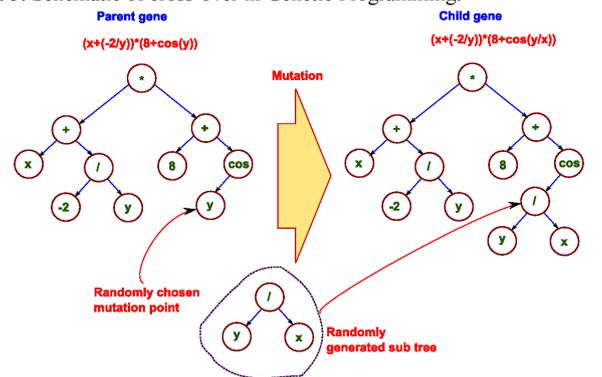
Fig. 6. Schematic of mutation in Genetic Programming.

In GP each candidate solution is a function itself and is encoded in the form of a tree. Fig. 5 shows the schematic for crossover between the two parent genes. Since the whole node with its corresponding sub-nodes get replaced in this case, so the crossover procedure is more effective and can provide a wide variety of individuals. Care must be taken so that the crossover process does not produce an indeterminate function or ill conditioned expression (e.g. division by zero, logarithm of a negative number etc.) and such solutions must be eliminated [22]. Fig. 6 shows the mutation schematic where a randomly chosen node in the tree is replaced by another randomly generated sub-tree giving rise to a new individual.

For the present study the population size is chosen to be 500. A tournament selection method is adopted and the tournament size is kept as 3. The maximum depth of each tree is assumed to be 7. Fig. 7 shows the Pareto front for the fitness value versus the number of terms of the expression found from GP where each dot represents a solution expression with different level of complexity and fitness value. The blue colors indicate the non-Pareto optimal solutions and the green colors indicate the Pareto optimal front. The solution having the lowest fitness is encircled by a red outline. It is obvious that the increase in the number of terms increases the complexity of the overall expression, but gives a better fit, i.e. a lower value of fitness function. However for ease of computability a trade-off is made between the fitness and complexity by intuitive judgment.

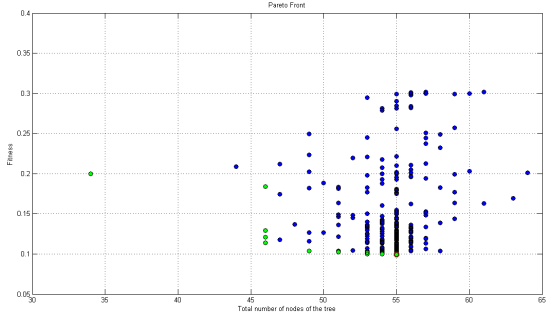

Fig. 7. Pareto front showing fitness vs. complexity of the genes.

Equation (13) shows the expressions for an optimal PID controller parameters obtained by the GP based symbolic regression method for normalized process gain ($K = 1$).

$$K_p = 1.033 + 0.1687 \times \left\{ \begin{array}{l} \sqrt{\left|\dfrac{\tau_{max}}{\cos(\tau_{min})}\right|} + \tanh\left(-\tau_{max}^2 + \dfrac{\tau_{min}}{\tau_{max}}\right) - \sqrt{L} \\ -\sin(\tau_{min}) - \sin\left(\begin{array}{l} 1.6 \times 10^{-6}\tau_{max}^2(1250L+2117) \\ \times\left(500\sqrt{\dfrac{L}{\tau_{max}}}\right) + 1877 - 500\dfrac{L}{\tau_{min}} \end{array}\right) \end{array} \right\}$$

$$K_i = 1.003 - 0.2452\sqrt{\left|\begin{array}{l} 4\ln(|\tau_{max}+L|) + 2\tanh(\tau_{min}) \\ + 3\tanh(L) + \tanh(\tau_{max}) - 0.8913 \end{array}\right|}$$

$$K_d = 1.399 + 0.09693 \times \left\{ \begin{array}{l} \sqrt{\left|\dfrac{\tau_{max}}{\cos(\tau_{min})}\right|} + \tanh\left(-\tau_{max}^2 + \dfrac{\tau_{min}}{\tau_{max}}\right) - \sqrt{L} \\ +\sqrt[4]{\tau_{max}} - \sin\left(\begin{array}{l} 1.6 \times 10^{-6}\tau_{max}^2(1250L+2117) \\ \times\left(500\sqrt{\dfrac{L}{\tau_{max}}}\right) + 1877 - 500\dfrac{L}{\tau_{min}} \end{array}\right) \\ -\dfrac{L\tau_{max}}{\tau_{min}} - \sin(\tau_{max}) + \tanh(-L+\tau_{min}) \\ +\cos\left(\dfrac{L\tau_{max}}{\tau_{min}}\right) - \sin(\tau_{min}) - \dfrac{6.457 \times 10^{-6}}{\tau_{max}} \end{array} \right\} \quad (13)$$

The optimal FOPID tuning rule for $K = 1$ is given by (14).

$$K_p = 1.1718 + 0.2726 \times \left\{ \begin{array}{l} -\dfrac{L}{\tau_{min}} - \tanh\left(L^2 + \dfrac{L}{\tau_{min}\tau_{max}} + \dfrac{\cos\left(\dfrac{\tau_{max}}{\tau_{min}}\right)}{e^{\tau_{max}}}\right) \\ -\dfrac{\sqrt{\left|\tanh\left(\dfrac{L}{\tau_{min}}\right) - \tau_{min}\right|}}{\left(\dfrac{\ln\dfrac{L}{\tau_{min}}}{\tau_{max}} + 2\tau_{max}\right)\left(\ln^2(\tau_{max}) + \dfrac{L}{\tau_{min}\tau_{max}^3} + \dfrac{L}{\tau_{min}}\right)} \end{array} \right\} \quad (14)$$

$$K_i = 0.3548 + 0.0783 \times \left\{ \begin{array}{l} -\dfrac{\left(\dfrac{\tau_{max}}{\tau_{min}}\right)^4}{\tau_{max}}\left(\dfrac{\tanh(\ln(\tau_{max}))}{0.503953^2}\right)^2 \\ +\left(\dfrac{\log(|0.1851\sin(\tau_{min})|)}{\tau_{max}}\right) \end{array} \right\}$$

$$K_d = 0.1379 + 0.1248 \times \left\{ \begin{array}{l} 1.6 \times 10^{-5}\left((250\tau_{min}-493)\sin(\sin(\tau_{max}))\right)^2 + \cos\left(\dfrac{\tau_{max}}{\tau_{min}}\right) \\ +\sqrt{\left|\sin\left(\dfrac{0.4861289}{L}\right)\right|} - \sin(\tau_{max}) + \sqrt{\left|\sin\left(\dfrac{\tau_{max}}{\tau_{min}}\right)\right|} + \sqrt{|2\tau_{min}|} \end{array} \right\}$$

$$\lambda = 0.9974 - 0.002605\left\{\sqrt{\tau_{max}L}\left(\tau_{max} - \tanh(\tau_{min})\right)\right\}$$

$$\mu = 2.0205 + 1.708 \times \left\{ \begin{array}{l} \tanh(\tanh(\tanh(L))) - \cos\left(\tanh\left(\dfrac{\tau_{max}}{\tau_{min}}\right)\right) \\ -\cos\left(\cos\left(\tanh\left(\dfrac{L\tau_{max}}{\tau_{min}}\right)\right)\right) + \cos\left(\cos\left(\dfrac{L\tau_{min}}{\tau_{max}^2 e^{\tau_{min}}}\right)\right) \\ -\cos\left(\tanh\left(L + \dfrac{L}{\tau_{max}} + \left(\dfrac{\tau_{max}}{\tau_{min}}\right)^2\right)\right) \end{array} \right\}$$

## C. Performance of the Tuning Rules

Four representative processes have been chosen from the four different classes of higher order process to validate the PID/FOPID tuning formula obtained by GP. Also, the GA based optimum control performances are compared with the rule based PID/FOPID controller, to show the wide applicability of such rules in process controls. Figs. 8-9 shows that the GP based analytical tuning rules closely follows the GA based optimal controller performances.

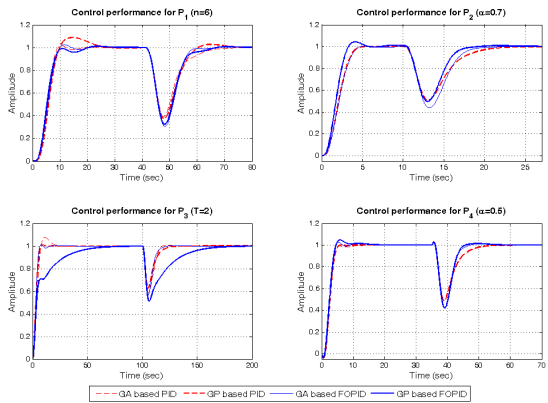

Fig. 8. Performance of the optimum PID/FOPID tuning rules.

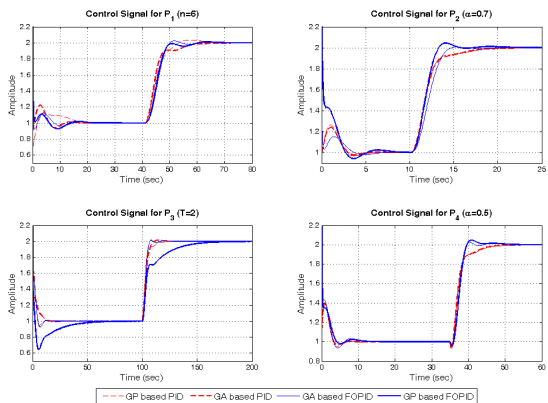

Fig. 9. Control signals with optimum PID/FOPID tuning rules.

## IV. CONCLUSION

Nyquist based new suboptimal model reduction technique via GA has been proposed in this paper which outperforms the existing $H_2$ norm based model reduction technique [18]. Few test-bench processes are modeled in SOPTD template using this technique and tuning rules for PID and FOPID controllers are extracted via symbolic regression using Genetic Programming. The rules are in the form of analytical expressions and hence are valuable to process control engineers due to ease of calculation and online implementation. The rules are chosen manually from the final set of GP outputs so that the complexity does not increase to a great extent and at the same time the accuracy is not compromised. The performance of the optimum tuning rules is demonstrated vis-à-vis the original GA based controller parameters, indicating nominal deterioration in the closed loop response of the control system.